\documentclass[manuscript,twocolumn]{aastex62}
\usepackage{txfonts}
\usepackage{latexsym,bm}
\usepackage{url}
\usepackage{color}
\usepackage{colortbl}
\usepackage{multirow}
\usepackage{ulem} 
\newcommand{\dee}{\mathrm{d}}
\definecolor{mygray}{gray}{.9}

\shorttitle{electron acceleration at shocks}

\shortauthors{Qin et al.}

\begin{document}

\title{Effects of shock and turbulence properties on electron acceleration} 
%at collisionless shocks}

\correspondingauthor{G. Qin}
\email{qingang@hit.edu.cn}

\author[0000-0002-3437-3716]{G. Qin}
\affiliation{School of Science, Harbin Institute of Technology, Shenzhen,
518055, China; qingang@hit.edu.cn}
%\affiliation{State Key Laboratory of Space Weather, National
%Space Science Center, Chinese Academy of Sciences,
%Beijing 100190, China}
%\affiliation{College of Earth Sciences, University of
%Chinese Academy of Sciences, Beijing 100049, China}

\author{F.-J. Kong}
\affiliation{State Key Laboratory of Space Weather, National
Space Science Center, Chinese Academy of Sciences,
Beijing 100190, China}
\affiliation{College of Earth Sciences, University of
Chinese Academy of Sciences, Beijing 100049, China}

\author{L.-H. Zhang}
\affiliation{State Key Laboratory of Space Weather, National
Space Science Center, Chinese Academy of Sciences,
Beijing 100190, China}
\affiliation{National Astronomical Observatories, Chinese Academy of Sciences;
Key Laboratory of Solar Activity, Chinese Academy of Sciences}

\begin{abstract}
Using test particle simulations we study electron acceleration at 
collisionless shocks with a two-component model turbulent magnetic 
field with slab component including dissipation range. We investigate
the importance of shock normal angle $\theta_{\text{Bn}}$, magnetic 
turbulence level $\left(b/B_0\right)^2$, and shock thickness on the 
acceleration efficiency of electrons. It is shown that at perpendicular 
shocks the electron acceleration efficiency is enhanced with the 
decreasing of $\left(b/B_0\right)^2$, and at $\left(b/B_0\right)^2=0.01$ 
the acceleration becomes significant due to strong drift electric field 
with long time particles staying near the shock front for shock drift 
acceleration (SDA). In addition, at parallel shocks the electron 
acceleration efficiency is increasing with the increasing of 
$\left(b/B_0\right)^2$, and at $\left(b/B_0\right)^2=10.0$ the 
acceleration is very strong due to sufficient pitch-angle scattering 
for first-order Fermi acceleration, as well as due to large local 
component of magnetic field perpendicular to shock normal angle 
for SDA.  On the other hand, the high perpendicular shock acceleration 
with $\left(b/B_0\right)^2=0.01$ is stronger than the high parallel shock 
acceleration with $\left(b/B_0\right)^2=10.0$, the reason might be the 
assumption that SDA is more efficient than first-order Fermi acceleration. 
Furthermore, for oblique shocks, the acceleration efficiency is small 
no matter the turbulence level is low or high. Moreover, for the effect 
of shock thickness on electron acceleration at perpendicular shocks, 
we show that there exists the bend-over thickness, $L_{\text{diff,b}}$. 
The acceleration efficiency does not change evidently if the shock 
thickness is much smaller than $L_{\text{diff,b}}$. However, if the 
shock thickness is much larger than $L_{\text{diff,b}}$, the 
acceleration efficiency starts to drop abruptly. 

\end{abstract}

\keywords{acceleration of electrons --- shock waves --- shock thickness}

\section{INTRODUCTION}

The spectra of energetic charged particles in astrophysical plasmas are 
usually in power law mainly generated by the acceleration at collisionless 
shock waves. One important physical mechanism for shock acceleration is 
the first-order Fermi acceleration \citep{Fermi1949,
Krymsky1977DoSSR...234..1306K,Axford1977ICRC...11..132A,
Bell1978MNRAS...182..147B,Blandford1978ApJ...221..L29B} in which charged 
particles gain energies by elastic scattering under magnetic fluctuations
across the shock. Another important physical mechanism is shock drift 
acceleration (SDA) \citep{Jokipii1982ApJ...255..716J, Forman1985, Lee1996, 
Shapiro2003, GuoXY2014}. In SDA, with non-zero background magnetic field 
perpendicular to the shock normal, if particles gyro-rotate near the shock 
plane with part of gyro-cycles in the upstream and the rest part in the 
downstream, they would get acceleration in each rotation cycle because of 
the drift electric field and different gyro-radii between the upstream and 
downstream regions. The first-order Fermi acceleration and shock drift 
acceleration are incorporated into the theory of diffusive shock 
acceleration (DSA). This theory predicts a power-law distribution downstream 
of the shock, so DSA is widely accepted as the source of astrophysical energetic 
particles. However, there are spacecraft observed spectra that are not in 
agreement with DSA theory such as an exponential-like rollover at higher 
energies \citep{Ellison1985}.

The shock acceleration of ions has been widely studied in many acceleration 
sites such as interplanetary traveling shocks, coronal shocks, Earth's bow shock, 
and the heliospheric termination shock in the past
\citep[e.g.,][]{Decker1986ApJ...306..710D,Decker1986JGR...91..13349D,
Desai2008JGR...113..A00B06D, LeRoux2009ApJ...693..534L,Florinski2009SSR...143..111F, 
NeergaardParker2012ApJ...757..97P, NeergaardParker2014ApJ...782..52P, KongEA17}.
In comparison with ions which usually have large gyro-radii, the low-energy 
electrons are thought to be difficult to interact with ambient magnetic
fluctuations due to their small gyro-radii $r_L$ which makes electrons 
resonant frequency high enough to be in the turbulence dissipation range. 
Therefore, there is a challenge for us to understand electron acceleration. 
On the one hand, space energetic electrons can be produced by
coronal shock waves associated with flares or coronal mass ejections (CMEs)
\citep{Uchida1973SP...28..495U,Vrsnak1995SP...158..331V,
Stewart1974SP...36..219S,Stewart1974SP...36..203S,Classen2002,Lara2003},
on the other hand, they can also be produced by Jovian magnetosphere 
\citep[e.g.,][]{Eraker1982,Moses1987}.
One direct evidence for electron acceleration is from solar type
$\mathrm{\uppercase\expandafter{\romannumeral2}}$ radio bursts, which are 
the radio signature of traveling CME-driven shocks in the solar corona
\citep{Klassen2002AA...385..1078K}. In addition, hard X-ray and $\gamma$-ray 
emission from impulsive solar flares \citep[e.g.,][]{Rieger1994ApJSS...90..645R}
demonstrates the production of energetic electrons by shock waves.
Based on the observations, there have been many theoretical studies aiming 
at explaining electron acceleration at coronal shock waves over the last four 
decades. \citet{Holman1983ApJ...267..837H} considered electron acceleration 
through a shock drift process and found that the production of type 
$\mathrm{\uppercase\expandafter{\romannumeral2}}$ emission was related 
to a high angle between the shock normal and the upstream magnetic field.
\citet{Tsuneta1998ApJ...495..L67T} studied first-order Fermi acceleration 
for non-thermal electrons produced in solar flares so that the impulsive 
hard X-ray source could be explained. \citet{Mann2001JGR...106..25323M}
showed that using a mirror (i.e., diffusive) acceleration mechanism at 
quasi-parallel shocks highly energetic electrons can be produced by 
shock waves in the solar corona.

In principle, the level of magnetic fluctuations has important effects 
on the shock acceleration, which should be different for shocks with 
varying obliquity. Low-energy electrons were found to interact with 
whistler waves \citep[e.g.,][]{Miller1996ApJ...461..445M}, which scatter 
electrons in pitch angle and enable them to diffuse in multiple crossings 
of the shock. To reveal the role whistler waves play in the acceleration process, 
numerical simulations using different turbulence models have been performed.
\citet{Giacalone2005ApJ...624..765G} showed that in the case of weak magnetic 
fluctuations the acceleration rate at parallel shocks is very small compared 
to perpendicular shocks, and that in the large-scale magnetic field fluctuations 
parallel shocks can efficiently accelerate particles to high energies. 
More recent numerical simulations of electron acceleration at shocks by
\citet{Guo2015ApJ...802..97G} studying shocks propagating through a 
kinematically defined turbulent magnetic field showed that with a significant 
turbulent variance electrons can be accelerated to high energies regardless of 
the angle between the shock normal and the upstream magnetic field. In their 
paper the authors studied electron acceleration for various shock-normal 
angles, and found that the acceleration efficiency is strongly dependent on 
the shock normal angle under weak magnetic fluctuations, but has a weak 
dependence on the shock normal angle for the case of large turbulence. 
Besides, it is also found that the shocks with higher angles accelerate 
electrons more efficiently than the ones with smaller shock angles, and that 
the energy spectrum does not notably depend on the average shock-normal angle 
when the magnetic fluctuations are sufficiently large. It is noted that in 
the work of \citet{Guo2015ApJ...802..97G}, dissipation range was not included 
in the turbulence. \citet{Li2013ApJ...769..22L} studied shock acceleration of
electrons by considering a turbulence model with power spectrum described by 
an inertial range and a dissipation range, to find the process in which 
electrons with lower energy gain energies through resonating in the dissipation
range of magnetic turbulence, so that energy spectra with high energy 
hardening were obtained. It is essential to study the dependence of the 
acceleration efficiency of electrons on the shock-normal angle within 
such a more realistic turbulence model 
including both inertial and dissipation ranges.

Shock thickness is another factor to affect the shock acceleration of electrons.
The observed shock transition layer can be very complicated: the magnetic profile 
only has ramp structures for low $\beta$ quasi-perpendicular shocks, while it may 
include foot, ramp, overshoot, and undershoot for high $\beta$, supercritical, 
quasi-perpendicular shocks \citep{Scudder1986JGR...91..11019S}.
\citet{Russell1982GRL...9..1171R} examined quasi-perpendicular shocks with low 
Mach numbers and plasma beta using ISEE-1 and ISEE-2 observations to find the 
shock thickness to be close to one ion inertial length ($\sim$$c/\omega_{pi}$).
\citet{Newbury1998JGR...103..29581N} showed that the width of the ramp transition 
at quasi-perpendicular, high Mach number shocks is within the range 0.5--1.5 
$c/\omega_{pi}$. \citet{Newbury1996GRL...23..781N} and \citet{Yang2013PP...20..092116Y},
however, reported that the ramp region of a very thin shock tends to be only 
a few electron inertial length $c/\omega_{pe}$. Moreover, statistical studies 
of the shock density transition by \citet{Bale2003PRL...91..265004B} concluded 
that the shock ramp scale is given by the convected ion gyroradius 
$v_{sh,n}/\Omega_{ci,2}$ over the range of Mach numbers 1--15, where $v_{sh,n}$ 
and $\Omega_{ci,2}$ indicate the shock velocity in the plasma frame and 
ion cyclotron frequency in the downstream of the shock, respectively.
Based on the above studies, numerically investigating of the impact of 
shock thickness on the acceleration efficiency is necessary for a better 
understanding of electron acceleration at collisionless shocks.

Based on \citet{KongEA17}, we use test particle simulations that include 
pre-existing magnetic field turbulence with two-component model 
\citep{QinEA02GRL, QinEA02APJ, Qin2002PhDT...1Q} 
to study acceleration of electrons at shocks. In this work, we include the 
dissipation range in the magnetic turbulence upstream and downstream of the 
shock because the resonant frequency of electrons in the turbulence is higher 
than that of ions due to the light mass of electrons. The layout of the paper 
is as follows. In Section 2, we describe the details of our numerical model. 
We show the results of the simulations in Section 3. 
In Section 4 we present our conclusions and discussion.

\section{MODEL}

The acceleration of particles at collisionless shocks is studied using
test particle simulations that include pre-existing electric and magnetic 
fields in the shock region. This approach is efficient in accelerating
the low-rigidity charged particles by describing their gyro-motions near 
the shock, and has been applied in many previous works 
\citep[][etc.]{Decker1986ApJ...306..710D, Decker1986JGR...91..13349D,
Giacalone2005ApJ...624..765G,Giacalone2009ApJ...701..1865G, KongEA17}.
Here, we use the model adopted from \citet{ZhangEA17} and \citet{KongEA17}. 

Firstly we illustrate the shock geometry. For simplicity we assume a 
hyperbolic function to describe a planar shock, which is similar to 
the fitted density transition in \citet{Bale2003PRL...91..265004B} 
and the flow speed model in \citet{Giacalone2009ApJ...701..1865G}. 
In the shock transition the plasma speed is given by
\begin{equation}
U(z)=\frac{U_1}{2s}\left\{(s+1)+(s-1)\tanh\left[\tan\left(-\pi z/L_{\text{diff}}
\right)\right]\right\}, \label{eq:uz}
\end{equation}
where $z=0$ is the location of the shock front, $z>0$ indicates plasma flow 
direction, 
$L_{\text{diff}}$ is the thickness of the shock transition. The shock parameters 
used in this study are listed in Table \ref{inputpara}: the upstream plasma speed 
is set as $U_1$ = 500 km s$^{-1}$ and the compression ratio $s=U_1/U_2$, along with 
pre-specified upstream mean magnetic field $B_{01}$ and Alfv$\acute{\text{e}}$n 
Mach number $M_{\text{A1}}$. Generally, we set $s=4$ for simplicity, 
but we also do some calculations with $s=2.6$ for comparison. 
We take the shock thickness $L_{\text{diff}}$ to be 9.28$\times10^{-6}$ AU 
if not otherwise stated. These parameters are in close analog with the values 
assumed by \citet{Guo2015ApJ...802..97G} for high-Mach number shocks, so that 
it is convenient to investigate the effect of different shock models on 
electron acceleration.

Note that in \citet{Guo2015ApJ...802..97G} the magnetic field was generated from the
magnetic induction equation with an isotropic turbulence spectrum, 
while in this work we employ a ``slab+2D" turbulence model as presented below.
The magnetic field is taken to be time independent %and spatially homogeneous,
with the form
\begin{equation}
\boldsymbol{B}(x,y,z)=\boldsymbol B_0(z)+\boldsymbol{b}(x,y,z),
\end{equation}
where $\boldsymbol B_0$ is the constant background field lying in the
$x-z$ plane and $\boldsymbol{b}$ is a zero-mean random magnetostatic turbulent
magnetic field transverse to $\boldsymbol B_0$. MHD Rankine--Hugoniot (RH)
conditions are satisfied on average. The turbulent field
$\boldsymbol{b}$ is composed of a slab component and a two-dimensional (2-D) 
component \citep{Mattaeus1990JGR...95..20673M, 
Zank1992JGR...97..17189Z, Bieber1996JGR...101..2511B, Gray1996GRL...23..965G, 
Zank2006JGR...111..A06108Z}. By following previous works \citep[e.g.,][]
{QinEA02APJ, BieberEA04, MatthaeusEA03} we assume
the slab correlation scale $\lambda$ = 0.02 AU. According to the recent 
research \citep{OsmanAHorbury07, DoschEA13, WeygandEA09, WeygandEA11}
we set the 2D correlation scale $\lambda_x=\lambda/2.6$, see also 
\citet{ShenAQin18}.

It is known that the turbulence spectrum in the solar wind consists of 
inertial range and dissipation range. In \citet{Li2013ApJ...769..22L}, 
the dissipation range in the turbulence spectrum (or called power spectrum 
density, PSD) was included to study the electron acceleration in solar flares. 
As the resonant wavenumber of low-energy electrons could be very large to 
lie in the dissipation range, we also include this range in the magnetic 
turbulence of the slab component in our model. The power spectrum of $P(k)$ 
is given by
\begin{equation}
	P(k)=C\frac{(1+k^{\prime2})^{-\beta_i/2}(k_b^{\prime2}+k^{\prime2})
	^{-\beta_d/2}}{(1+k_b^{\prime2}+k^{\prime2})^{-\beta_i/2}},
\end{equation}
where $\beta_i$ and $\beta_d$ are the spectral indices in the inertial
and dissipation ranges, respectively, $k_i=1/\lambda$, $k^\prime=k/k_i$,
and $k_b^\prime=k_b/k_i$. $k_i$ is the break wavenumber that separates 
from the energy range to inertial range, and $k_b$ is the break wavenumber
that separates from the inertial range to dissipation range.
We assume spectral index in the inertial range $\beta_i=5/3$ as Kolmogorov 
cascading. Spacecraft observations of the dissipation range index $\beta_d$
vary considerably \citep[e.g.,][]{Smith2006}, however, for simplicity, 
generally we set a constant $\beta_d=2.7$ according to 
\citet{Li2013ApJ...769..22L} \citep[see also,][]{Leamon1999JGR, Howes2008JGR, 
Chen2010PRL}, and we also do some calculations with $\beta_d=3.4$ for comparison. 
We take the value $k_b=3\times 10^4/\lambda=10^{-5}$ m$^{-1}$, which is equal to
the minimum wavenumber of $\sim$7.88 keV electrons and satisfies the requirement 
of \citet{Leamon1999JGR}. Regarding the magnetic energy density ratio of 
different turbulence component, although it can vary considerably in the 
dissipation range from spacecraft observations \citep[e.g.,][]{Oughton2015}, 
we assume $E_{slab}:E_{2D}=20:80$ for simplicity in the work. It should be also 
noted that from the resonance condition ($\omega-k_\parallel v_\parallel=n\Omega$, 
where $\omega$ and $\Omega$ are the wave frequency and electron cyclotron frequency, 
respectively), the resonant wave-vector is parallel to the background magnetic field, 
which corresponds to the slab component. In addition, to generate turbulence
spectrum extending to dissipation range much more computing resources are needed 
for 2D component than that for slab component. Therefore, we do not include the 
dissipation range in the magnetic turbulence of 2D component 
\citep{Qin2002PhDT...1Q, QinEA02APJ}.

\begin{figure}
\epsscale{1.} \plotone{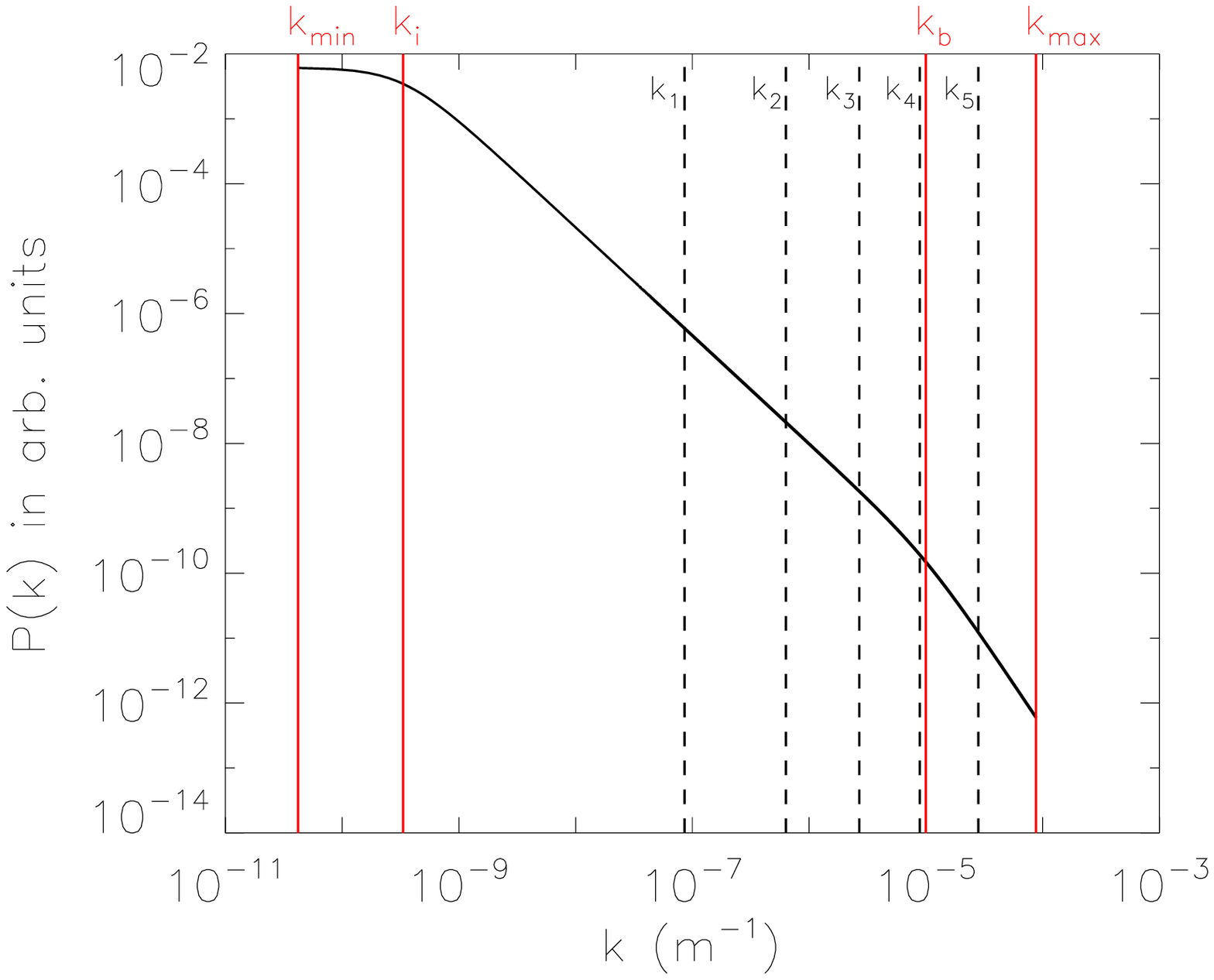}
\caption{$P(k)$ in arbitrary units as a function of $k$ with
the energy, inertial, and dissipation ranges. The wavenumber $k_{min}$, $k_i$,
$k_b$, and $k_{max}$ are indicated in red lines, and wavenumber $k_1$, $k_2$, 
$k_3$, $k_4$, and $k_5$ are indicated in dashed lines.
\label{Pk}}
\end{figure}

In addition, we describe some important parameters relevant to the numerical
simulation box. For the spatial domain size in $x$, $y$, and $z$ directions, we
take $x_{\text{box}}=y_{\text{box}}=z_{\text{box}}=10^4\lambda$, which defines 
a box large enough so that electrons are not easy to escape. The power spectrum 
of turbulence is continuous and nonperiodic in nature, so there is no way to 
realize the actual turbulence in the simulations. We here adopt a turbulence box 
of size $L_x=L_y=10\lambda$, and $L_z=50\lambda$, for the 2D and slab components, 
respectively. In fact, we have checked that it is very difficult for electrons 
to transport in space larger than the size of the slab-2D box. The number of grids 
in the slab modes is set as $N_z=2^{22}$, and the minimum and maximum wavenumber 
of the slab turbulence are $k_{min}=2\pi/L_z\sim4.2\times 10^{-11}$ m$^{-1}$ and
$k_{max}=2\pi N_z/2L_z\sim8.8\times 10^{-5}$ m$^{-1}$, respectively.
We plot the $P(k)$ in arbitrary units as a function of $k$ in Figure \ref{Pk},
with the wavenumber $k_{min}$, $k_i$, $k_b$, and $k_{max}$ indicated in red 
vertical lines. We also show the minimum resonant wavenumber $k_1$, $k_2$, $k_3$, 
$k_4$, and $k_5$ indicated in dashed lines for electrons with energies of $10,000$, 
$1,000$, $100$, $10$, and $1$ keV, respectively. It can be clearly seen that 
low-energy electrons which are below 10 keV resonate in the dissipation range, 
while the electrons with energies above 10 keV may resonate in the inertial range.
For the 2D modes, we set the number of grids as $N_x=N_y=4096$. Note that we
take smaller box size in perpendicular direction than that in parallel direction
since the movement range of particles in perpendicular directions is much
smaller than that in parallel direction because of the smaller perpendicular 
diffusion coefficients. For more details on the settings and realization of 
the slab-2D magnetic field model in numerical code, see \citet{QinEA02GRL,QinEA02APJ}, 
\citet{Qin2002PhDT...1Q}, \citet{ZhangEA17}, and \citet{KongEA17}.

A large number of electrons with energies of 1 keV in the upstream plasma frame 
at $z_0=-5.80\times 10^{-5}$ AU are isotropically injected, and each electron's 
trajectory is obtained by solving the Lorentz motion equation in the shock frame 
of reference with a fourth-order Runge--Kutta method with an adjustable time step
which maintains accuracy of the order of $10^{-9}$. The motion equation is 
given by
\begin{equation}
 \frac{\dee\boldsymbol{p}}{\dee t}=q\left[\boldsymbol{E}(\boldsymbol{r},t)
+\boldsymbol{v}\times\boldsymbol{B}(\boldsymbol{r},t)\right],
\end{equation}
where $\boldsymbol{p}$ is the particle momentum, $\boldsymbol{v}$ is the
particle velocity, $q$ is the electron charge, t is time, and the frame
of reference is moving with the shock front. 
$\boldsymbol{E}=-\boldsymbol{U}\times\boldsymbol{B}$ 
is the convective electric field under the MHD approximation.

\section{Numerical RESULTS}

\subsection{Effects of shock geometry and turbulence level}
Using numerical simulations, we first investigate the acceleration of 
electrons at shocks with different angles $\theta_{\text{Bn}}$ between 
the shock normal and the upstream mean magnetic field, and different 
levels of magnetic turbulence $\left(b/B_0\right)^2$. We change 
$\theta_{\text{Bn}}$ from 0$^\circ$ to 90$^\circ$ with 15$^\circ$ interval. 
For each value of $\theta_{\text{Bn}}$, $\left(b/B_0\right)^2$ varies in 
four values, $0.01$, $0.1$, $1.0$, and $10.0$. In each simulation with 
certain values of $\theta_{\text{Bn}}$ and $\left(b/B_0\right)^2$, 
we calculate the trajectories of electrons for $t_{\text{acc}}=23.2$ min, 
which is equivalent to the value of the acceleration time in 
\citet{Guo2015ApJ...802..97G}, and at the end of simulations the 
kinetic energy is computed in the reference frame of shock front.

\begin{figure}
\epsscale{1.} \plotone{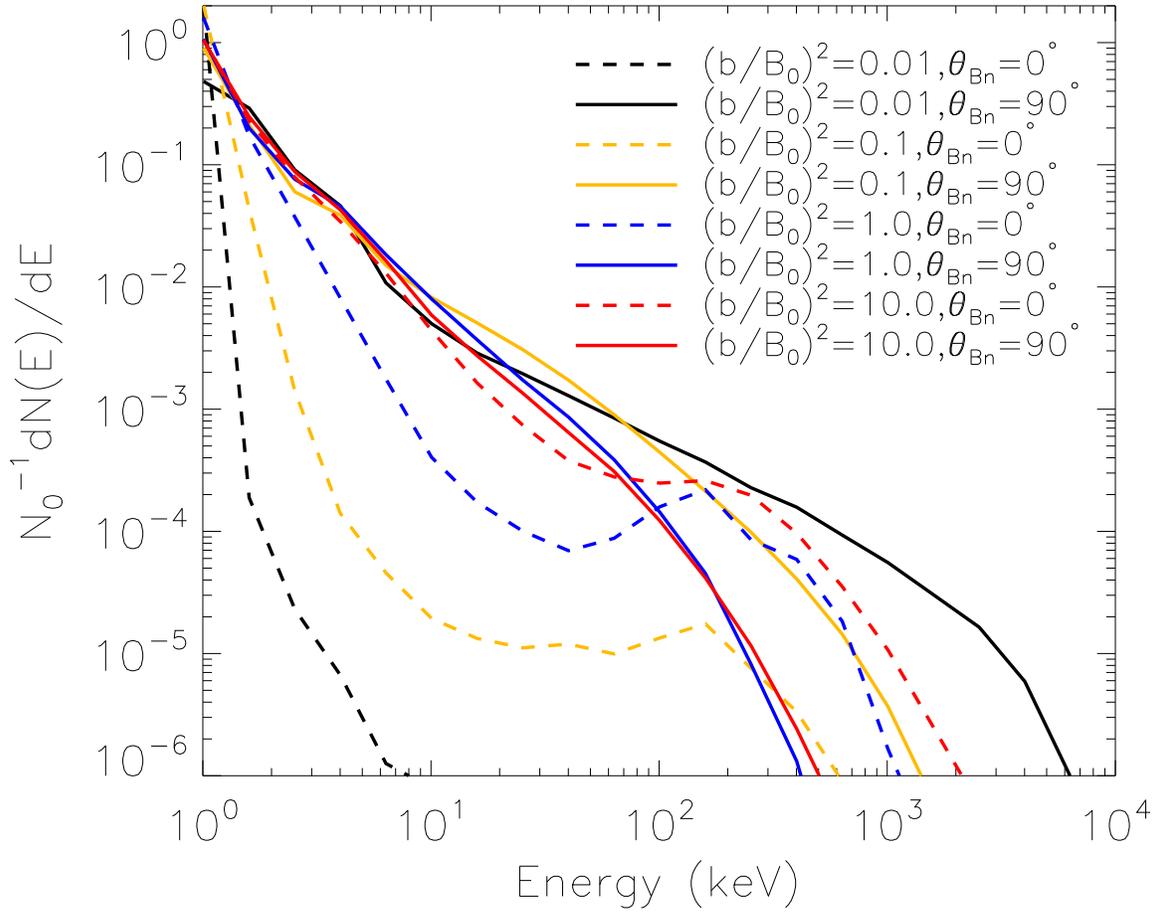}
\caption{Simulated downstream energy spectra at $t_{\text{acc}}=23.2$ min
for different shock-normal angles $\theta_{\text{Bn}}$ and turbulence levels 
$\left(b/B_0\right)^2$. Dashed and solid lines indicate $\theta_{\text{Bn}}=0^\circ$
and $\theta_{\text{Bn}}=90^\circ$, respectively. 
Black, yellow, blue, and red lines are for $\left(b/B_0\right)^2=0.01$, 0.1, 1.0, and
10.0, respectively. 
%The red dotted line is for the quasi-parallel
%shock with $\theta_{Bn}=30^\circ$ and $\left(b/B_0\right)^2=10.0$.
The kinetic energy of electrons is measured in the shock frame.\label{flux}}
\end{figure}

Figure \ref{flux} shows the energy spectra of electrons downstream of
the shock at the end of simulations for different shock normal angles 
$\theta_{\text{Bn}}$ and turbulence levels $\left(b/B_0\right)^2$. 
Dashed and solid lines indicate $\theta_{\text{Bn}}=0^\circ$ and 
$90^\circ$, respectively. Black, yellow, blue, and red lines are for 
$\left(b/B_0\right)^2=0.01$, $0.1$, $1.0$, and $10.0$, respectively. 
For the cases of parallel shocks (dashed lines) we have following results.
It is shown that with low turbulence levels, $\left(b/B_0\right)^2=0.01$ 
and $0.1$, the energy spectrum curves of black and yellow dashed lines, 
respectively, abruptly decrease with the increase of electron energy in 
the range around 1--10 keV. However, with $\left(b/B_0\right)^2=0.1$, 
the energy spectrum curve has higher level in the energy range around 
1--10 keV, and it becomes flat in the higher energy range around 
10--200 keV, and then decreases again in the even higher energy range 
around 200--600 keV. The energy spectrum with higher turbulence level, 
$\left(b/B_0\right)^2=1.0$, indicated by blue dashed line is larger than 
the one with $\left(b/B_0\right)^2=0.1$, and the spectral index in the 
range around 1--200 keV is also larger, and the energy of particles 
extends as high as $1$ MeV. When the turbulence level increases to 
$\left(b/B_0\right)^2=10.0$ (red dashed line), the energy spectrum 
becomes even larger, and electrons are found to be accelerated to the
energies as high as $2$ MeV. On the other hand, for perpendicular shocks 
($\theta_{\text{Bn}}=90^\circ$), all of the spectra curves of the 
accelerated particles downstream of the shocks for 
$\left(b/B_0\right)^2=0.01$, $0.1$, $1.0$ and $10.0$ indicated by black, 
yellow, blue, and red solid lines, respectively, are similar in the 
energy range around 1--10 keV, but in the energy range $\gtrsim 10$ keV, 
the ones with $\left(b/B_0\right)^2=0.01$ and $0.1$ are larger than that 
with $\left(b/B_0\right)^2=1$ and $10$, and in the range $\gtrsim 100$ keV, 
the one with $\left(b/B_0\right)^2=0.01$ is much larger than the one with 
$\left(b/B_0\right)^2=0.1$. In addition, the spectrum for parallel shock 
(parallel spectrum hereafter) with $\left(b/B_0\right)^2=10$ is similar 
as the spectra for perpendicular shocks (perpendicular spectra hereafter) 
in the energy range around 1--10 keV. Furthermore, the perpendicular one 
with $\left(b/B_0\right)^2=0.01$ is larger than the parallel one with 
$\left(b/B_0\right)^2=10$ in the energy range $\gtrsim 10$ keV, but the 
perpendicular one with $\left(b/B_0\right)^2=0.1$ is smaller than the 
parallel one with $\left(b/B_0\right)^2=10$ in the energy range 
$\gtrsim 200$ keV, and the perpendicular ones with $\left(b/B_0\right)^2=1$ 
and $10$ are smaller than the parallel ones with $\left(b/B_0\right)^2=1$ 
and $10$ in the energy range $\gtrsim 100$ keV.

\begin{figure}
\epsscale{1.} \plotone{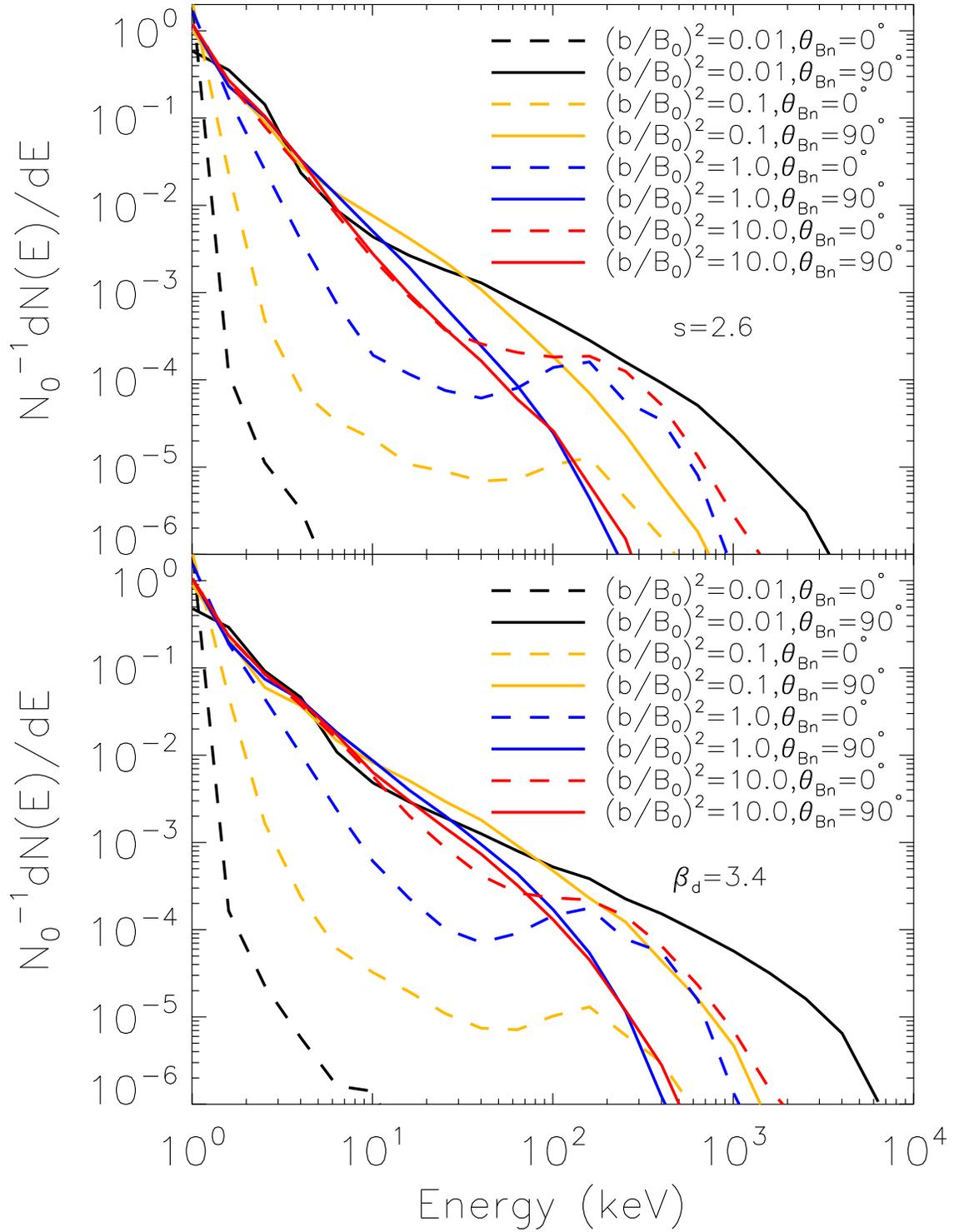}
\caption{Top and bottom panels are similar as Figure \ref{flux}, except that 
$s=2.6$ and $\beta_d=3.4$, respectively.\label{flux1}}
\end{figure}

In this work we generally do not vary compression ratio $s$ for
simplicity, and we consider strong shock acceleration, so we set $s=4$.
For comparison, we also make simulations with $s=2.6$. 
Moreover, because spacecraft observations of the spectral index in the
dissipation range vary in the range $-4<\beta_d<-1$
\citep[e.g.,][]{Smith2006}, we perform additional simulations with a
spectral index value $\beta_d=3.4$ for comparison.
Top and bottom panels of Figure \ref{flux1} show simulation results for 
cases similar as that in Figure \ref{flux} except that $s=2.6$ and 
$\beta_d=3.4$, respectively. In both panels of Figure \ref{flux1}, 
the spectral features for various values of the turbulence level at 
parallel and perpendicular shocks, in general, are similar to those 
shown in Figure \ref{flux}. For instance, in the energy range of 
1--10 keV, the spectra for perpendicular shocks and that for parallel 
shocks with $(b/B_0)^2=10.0$ are similar, with values higher than that 
for parallel shocks with $(b/B_0)^2=0.01$, $0.1$, and $1.0$. In addition, 
the perpendicular shocks for low turbulence levels ($(b/B_0)^2=0.01$ and $0.1$)
accelerate electrons to higher energies compared to that for high turbulence 
levels ($(b/B_0)^2=1.0$ and $10.0$).  The details of the energy spectra in 
Figure \ref{flux1}, however, are different from that shown in Figure 
\ref{flux}. For example, for weak shocks with $s=2.6$ in the top panel, in 
any condition of turbulence level and $\theta_{\text{Bn}}$, the spectra 
extend to lower energies in comparison with the strong shocks with $s=4.0$ 
shown in Figure \ref{flux}.  
In addition, in the energy range about 10--200 keV, for the case of steeper 
spectral slope in the turbulence dissipation range ($\beta_d=3.4$) in the bottom 
panel of Figure \ref{flux1}, the spectrum for $(b/B_0)^2=0.1$ at parallel shock is 
less flatter compared to the similar case but $\beta_d=2.7$ in Figure \ref{flux}. 
Figure \ref{flux1} suggests that $s=4$ and $\beta_d=2.7$ can be used as 
representative parameters to study shock acceleration.
Therefore, in the rest of the paper, we study electron acceleration at shocks
in fixed compression ratio, $s=4$, and spectral index in dissipation range 
$\beta_d=2.7$.

\begin{figure}
\epsscale{1.} \plotone{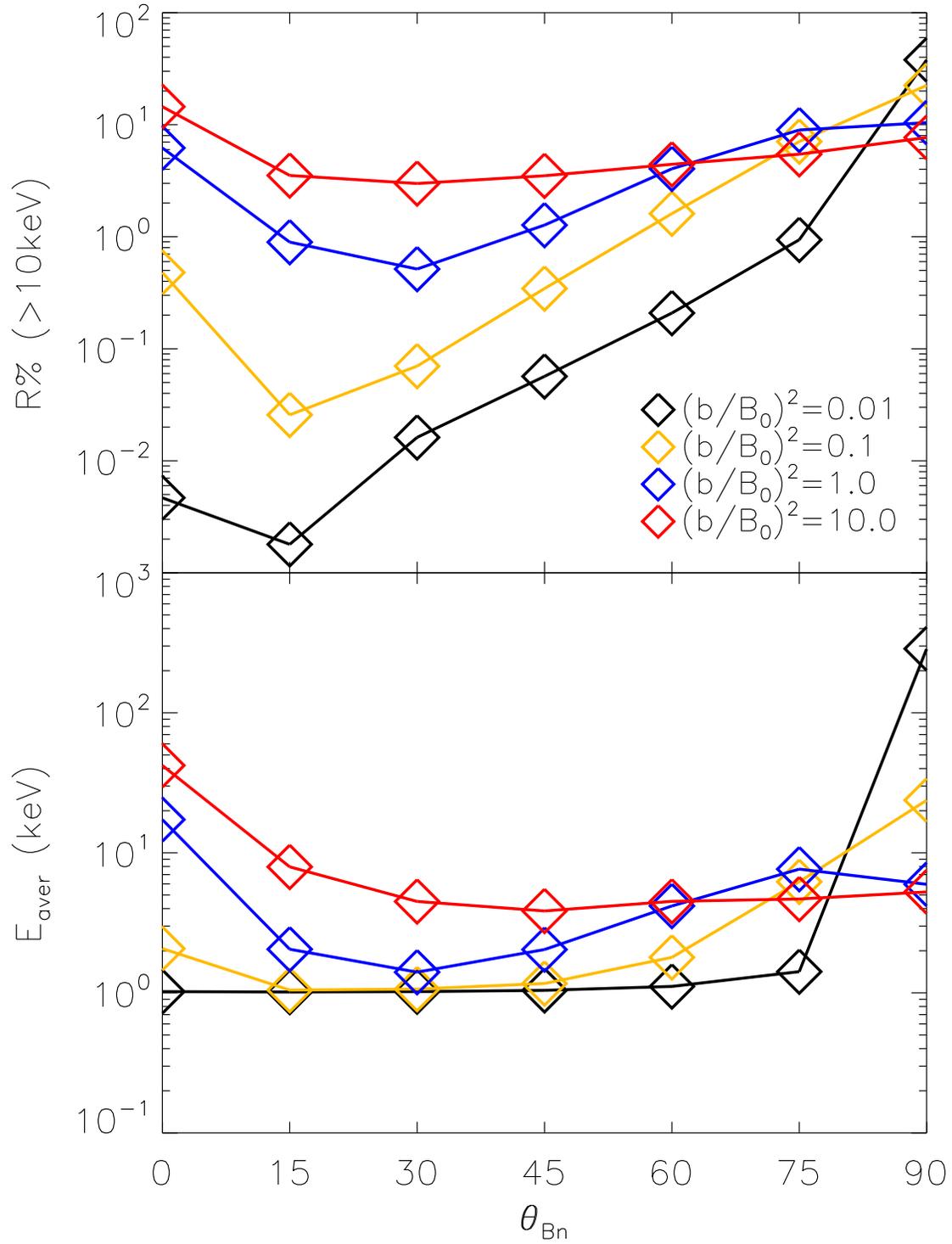}
\caption{The acceleration efficiency of electrons versus shock-normal angle
	for various turbulence levels. $R\%$ (top panel) and $E_{\text{aver}}$ 
	(bottom panel)
represents the fraction of accelerated electrons
with energies of more than 10 keV and average energy of electrons, respectively,
at the end of the simulations. \label{effi}}
\end{figure}

In Figure \ref{effi} as a function of shock normal angle $\theta_{\text{Bn}}$,
we present the percentage $R\%$ of electrons accelerated to more than 10 keV 
(top panel) and average electron energy $E_{\text{aver}}$ (bottom panel) 
at the end of simulations, for four different turbulence levels 
$\left(b/B_0\right)^2=0.01$, $0.1$, $1.0$, and $10$ indicated by black, 
yellow, blue, and red lines, respectively. Here, we use $R\%$ and 
$E_{\text{aver}}$ as measures of the efficiency of electron acceleration. 
It can be seen that the accelerated percentage, $R\%$, and average energy 
$E_{\text{aver}}$ have the same trend as a function of $\theta_{\text{Bn}}$, 
i.e., they decrease with $\theta_{\text{Bn}}$ increasing from $0^\circ$ to 
$15^\circ$, and then generally increase with $\theta_{\text{Bn}}$  increasing 
from $30^\circ$ to $90^\circ$. In addition, with $\theta_{\text{Bn}}\lesssim 
60^\circ$  ($\theta_{\text{Bn}}> 75^\circ$), $R\%$ and $E_{\text{aver}}$ 
increase (decrease) with the increasing of turbulence level. 
It is shown that, in parallel shocks with $\theta_{\text{Bn}}\sim 0^\circ$, 
$R\%$ and $E_{\text{aver}}$ are much larger with high turbulence level than 
that with low turbulence level, on the other hand, in perpendicular shocks 
with $\theta_{\text{Bn}}\sim 90^\circ$, $R\%$ and $E_{\text{aver}}$ are much 
larger with low turbulence level than that with high turbulence level. 
Furthermore, as $\theta_{\text{Bn}}\gtrsim30^\circ$ the acceleration 
efficiency in general increases with increasing of $\theta_{\text{Bn}}$,
and its largest variation with $\theta_{\text{Bn}}$ is seen at a low 
turbulence level $\left(b/B_0\right)^2=0.01$. $R\%$ and $E_{\text{aver}}$ 
in the case of perpendicular shock and $\left(b/B_0\right)^2=0.01$ are 
larger than that in the case of parallel shock and $\left(b/B_0\right)^2=10$. 
We can also see that in high turbulence level, the shock acceleration 
efficiency of electrons is in weak dependence of shock normal angle, 
especially when $\theta_{\text{Bn}}>0^\circ$. 
 
\begin{figure}
\epsscale{1.} \plotone{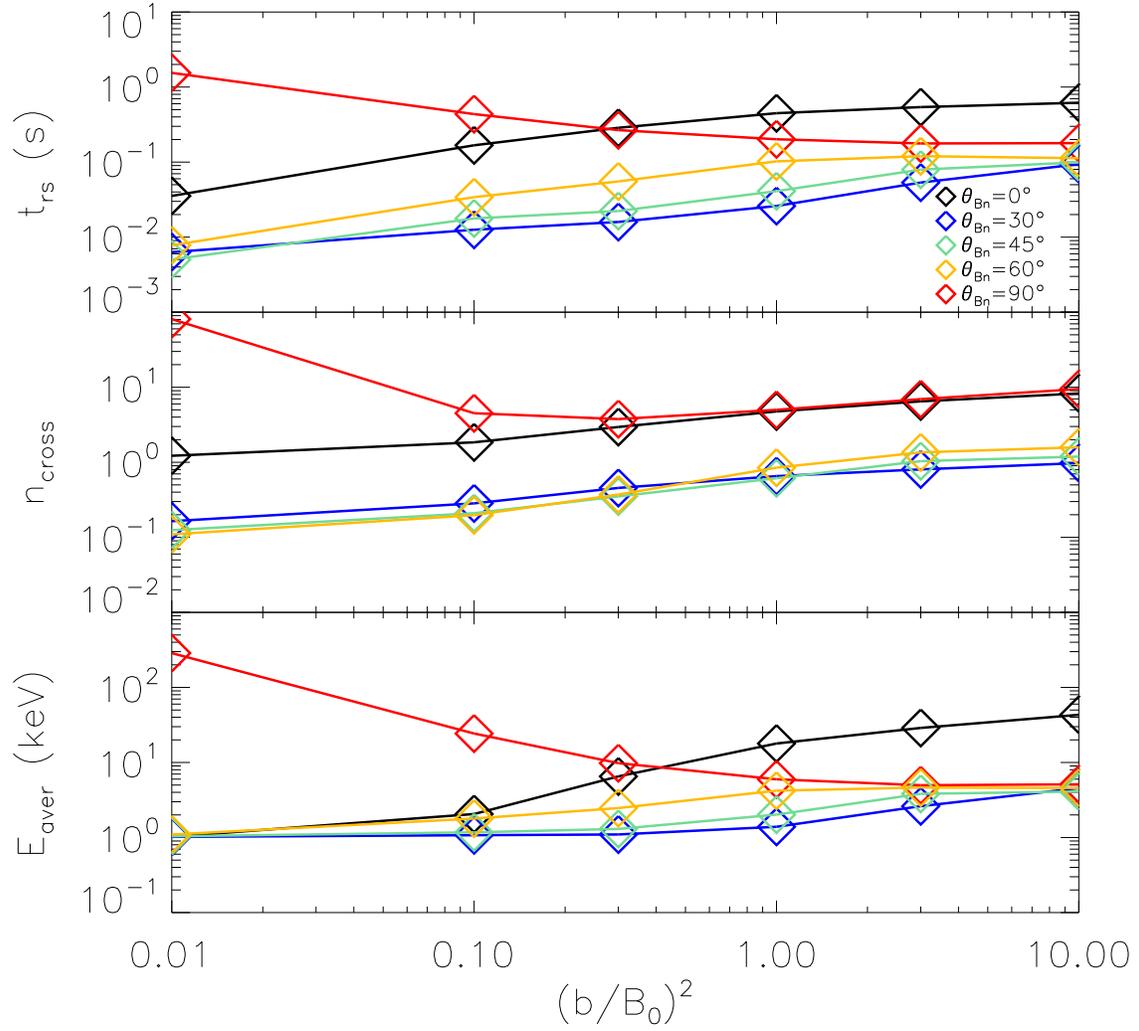}
\caption{Average time particles 
	staying within a gyration radius from the shock front $t_{\text{rs}}$,
%transition layer 
average shock crossing times $n_{\text{cross}}$, and average energy 
$E_{\text{aver}}$, in top, middle, and bottom panels, respectively, as a 
function of turbulence level. Black, blue, green, yellow,
and  red lines indicate $\theta_{\text{Bn}}=0^\circ$, $30^\circ$, $45^\circ$, 
$60^\circ$, $90^\circ$, respectively.
\label{time}}
\end{figure}

In Figure \ref{time} we continue to study shock acceleration of electrons.
Black, blue, green, yellow, and red lines indicate 
$\theta_{\text{Bn}}=0^\circ$, $30^\circ$, $45^\circ$, $60^\circ$, and $90^\circ$, 
respectively. The top panel shows average time particles staying within a 
gyration radius from the shock front $t_{\text{rs}}$ as a function of 
turbulence level $\left(b/B_0\right)^2$. From this panel it is shown that 
with low turbulence level, $\left(b/B_0\right)^2=0.01$, particles would stay 
within a gyration radius from the shock front for a long and short time in 
the cases of perpendicular and parallel shocks, respectively. In addition, 
for oblique shocks, i.e., $\theta_{\text{Bn}}=30^\circ$, $45^\circ$, and 
$60^\circ$, $t_{\text{rs}}$ is much smaller than that for both parallel 
and perpendicular shocks. As $\left(b/B_0\right)^2$ increases from $0.01$ 
to $10$, $t_{\text{rs}}$ for perpendicular shocks is decreasing while 
$t_{\text{rs}}$ for parallel and oblique shocks is increasing. With 
$\left(b/B_0\right)^2\sim 0.3$, $t_{\text{rs}}$ for perpendicular and 
parallel shocks are equal, and with $\left(b/B_0\right)^2\gg 0.3$, 
$t_{\text{rs}}$ for parallel shocks is larger than that for perpendicular ones. 
However, $t_{\text{rs}}$ for perpendicular shocks with $\left(b/B_0\right)^2=0.01$ 
is much larger than that for parallel ones, and $t_{\text{rs}}$ for oblique 
shocks is always smaller than that for perpendicular and parallel ones. 
The middle panel shows average crossing times $n_{\text{cross}}$ as a function 
of $\left(b/B_0\right)^2$. From this panel it is shown that for parallel shocks
particles would cross the shock front more times with the increasing of 
$\left(b/B_0\right)^2$. However, for perpendicular shock acceleration, 
$n_{\text{cross}}$ would decrease as $\left(b/B_0\right)^2$ increases from 
$0.01$ to $\sim 0.3$, and it would have the similar value as that for parallel
shocks with $\left(b/B_0\right)^2\gtrsim 0.3$. In addition, for oblique shocks, 
$n_{\text{cross}}$ has the similar trend as that for parallel shocks but in a much 
lower level. 
Bottom panel shows average energy of particles at the end of simulations 
$E_{\text{aver}}$ as a function of turbulence level. From this panel it 
is shown that as $\left(b/B_0\right)^2$ is very small, i.e., 
$\left(b/B_0\right)^2=0.01$, $E_{\text{aver}}\gg E_0$ for perpendicular 
shocks, but $E_{\text{aver}}\sim E_0$ for parallel and oblique shocks. 
As $\left(b/B_0\right)^2$ increases from $0.01$ to $10$, $E_{\text{aver}}$ 
decreases and increases for perpendicular and non-perpendicular shocks, 
respectively. However, $E_{\text{aver}}$ for parallel shocks would increases 
more sharply than that for oblique shocks with $\left(b/B_0\right)^2$ in the 
range from $0.1$ to $10$, so $E_{\text{aver}}$ for parallel shocks is much 
larger than that for oblique shocks with $\left(b/B_0\right)^2\gtrsim 0.1$. 
With $\left(b/B_0\right)^2\lesssim 0.4$ and $\gtrsim 0.4$, $E_{\text{aver}}$ 
for perpendicular shocks is larger and smaller than that for parallel ones, 
respectively.

From Figure \ref{time} we can see that as turbulence level is very low, for
perpendicular shocks particles could stay near the shock front for a long time, 
so they can get large acceleration from SDA, and for parallel shocks particles 
could not get efficient parallel scatterings to cross the shock front for 
many times to get large acceleration from first-order Fermi scattering process. 
As $\left(b/B_0\right)^2$ increases, for perpendicular shocks, the local magnetic 
fields would have larger component across the shock front so that it is more difficult 
for particles to keep near the shock front to get acceleration from SDA, but for 
parallel shocks, particles would get stronger parallel scatterings to cross the 
shock front many times and get acceleration from first-order Fermi process. 
In addition, for parallel shocks with large $\left(b/B_0\right)^2$, there is 
strong local magnetic field component perpendicular to the shock normal direction, 
so particles could also get acceleration from SDA. Furthermore, for oblique shocks, 
following results could be obtained: Firstly, particles can not stay near the shock 
front for long time to get large acceleration from SDA since there is strong 
cross shock magnetic component. Secondly, as $\left(b/B_0\right)^2$ is large, 
the parallel scatterings of particles only have partial contribution to shock 
front crossing because of the field obliquity, so that particles can not get 
large acceleration from first-order Fermi process too. Therefore, oblique shocks 
have low acceleration effects for both low and high turbulence levels.

\subsection{Effects of shock thickness}

We next examine the effect of shock thickness on electron acceleration 
by perpendicular shocks with $\left(b/B_0\right)^2=0.1$ and $1$, 
by varying $L_{\text{diff}}$ in Equation (\ref{eq:uz}).

\begin{figure}
\epsscale{1.} \plotone{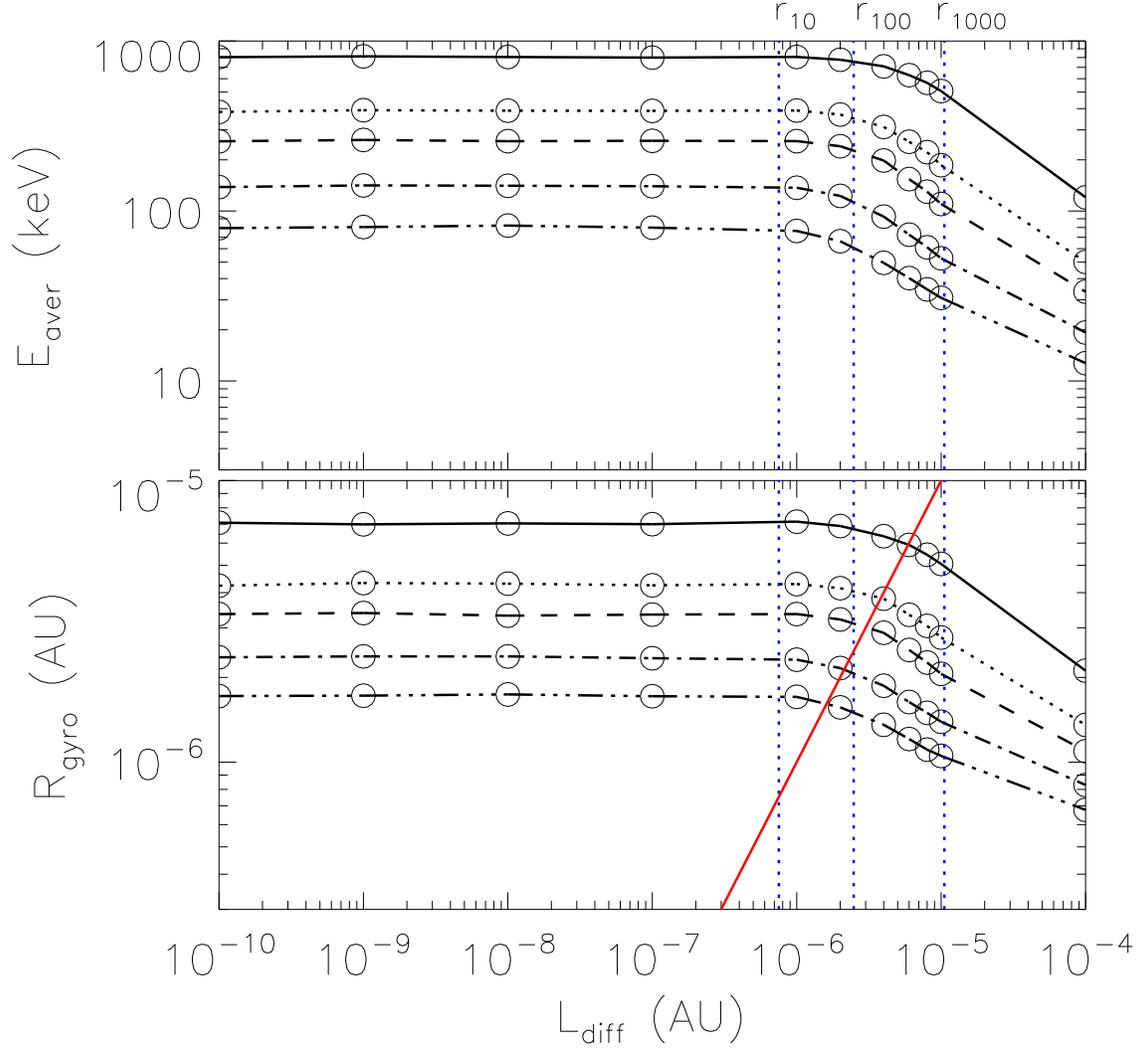}
\caption{Average energy (upper panel) and average gyro-radii considering
pitch angle and upstream background magnetic field (bottom panel) versus
shock thickness $L_{\text{diff}}$ for $\theta_{\text{Bn}}=90^\circ$ and 
$\left(b/B_0\right)^2=0.1$.
Solid, dotted, dashed, dash-dotted, and dash-dotted-dotted lines indicate
particles from 
top, second, third, fifth, and seventh $2\%$ of accelerated electrons ordered 
in energy.
	The dotted vertical lines in each panel indicate the 
	$L_{\text{diff}}$ equals to the
gyro-radii of 10, 100 and 1000 keV electrons.
Red line indicates $R_{\text{gyro}}=L_{\text{diff}}$.
\label{ldiff0}}
\end{figure}

Figure \ref{ldiff0} shows the average energy (top panel) and average 
gyro-radii considering pitch angle and upstream background magnetic 
field (bottom panel) versus shock thickness $L_{\text{diff}}$ for 
$\theta_{\text{Bn}}=90^\circ$ and $\left(b/B_0\right)^2=0.1$.
At the end of simulations with $t_{\text{acc}}=23.2$ min, we calculate 
average energy and gyro-radii of particles with significant acceleration, 
i.e., for particles from top, second, third, fifth, and seventh $2\%$ of 
accelerated electrons ordered in energy with solid, dotted, dashed, 
dash-dotted, and dash-dotted-dotted  lines, respectively. The blue dotted 
vertical lines (from left to right) indicate the $L_{\text{diff}}$ as the 
maximum gyro-radii of $10$ keV ($r_{10\text{ keV}}\sim 1.13\times 10^5$ m), 
$100$ keV ($r_{100\text{ keV}}\sim3.72\times 10^5$ m), and $1000$ keV 
($r_{1000\text{ keV}}\sim1.58\times 10^6$ m) electrons in the upstream. 
In the bottom panel, the red solid line indicates $R_{\text{gyro}}=L_{\text{diff}}$.
It is shown that there exists a bend-over thickness $L_{\text{diff,b}}$, 
the average energy does not have obvious variation for thin shock thickness 
$L_{\text{diff}}\lesssim L_{\text{diff,b}}$, while it decreases rapidly with 
the increasing of $L_{\text{diff}}$ when $L_{\text{diff}}\gtrsim L_{\text{diff,b}}$. 
It is seen that $L_{\text{diff,b}}$ is in the scale of the average gyro-radii of particles.

\begin{figure}
\epsscale{1.} \plotone{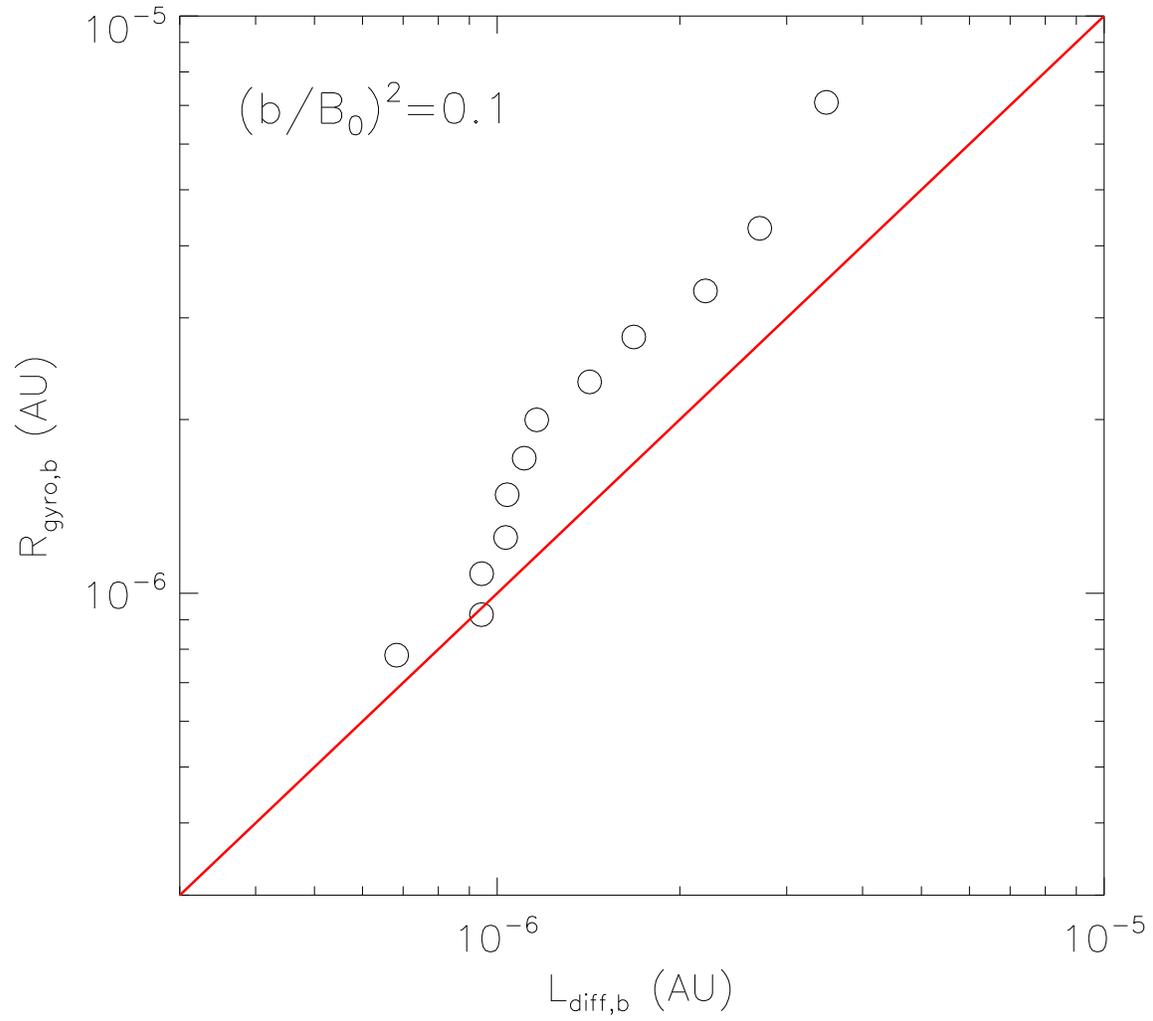}
	\caption{$R_{\text{gyro,b}}$ as a function of $L_{\text{diff,b}}$ with 
	$\left(b/B_0\right)^2=0.1$. Solid line indicates 
    $R_{\text{gyro,b}}=L_{\text{diff,b}}$.
\label{ldiff0b}}
\end{figure}

For each curve of $R_{\text{gyro}}$ as function of $L_{\text{diff}}$ 
of the top, second, third, ..., twelfth $2\%$ of accelerated particles 
ordered in energy, we fit the four data points of the smallest $L_{\text{diff}}$ 
as a line in log-log space, and fit the five data points of the largest 
$L_{\text{diff}}$ as another line in log-log space. The intersection of 
the two lines are assumed to be the bend-over point, denoted as $(L_{\text{diff,b}},
R_{\text{gyro,b}})$. Figure \ref{ldiff0b} shows $R_{\text{gyro,b}}$ 
as a function of $L_{\text{diff,b}}$ from the fitting, with the red 
line indicating $R_{\text{gyro,b}}=L_{\text{diff,b}}$. It is shown 
that generally the bend-over points of the curves, 
$E_{\text{aver}}$--$L_{\text{diff}}$, can be approximately expressed 
as $R_{\text{gyro,b}}=L_{\text{diff,b}}$. 

\begin{figure}
\epsscale{1.} \plotone{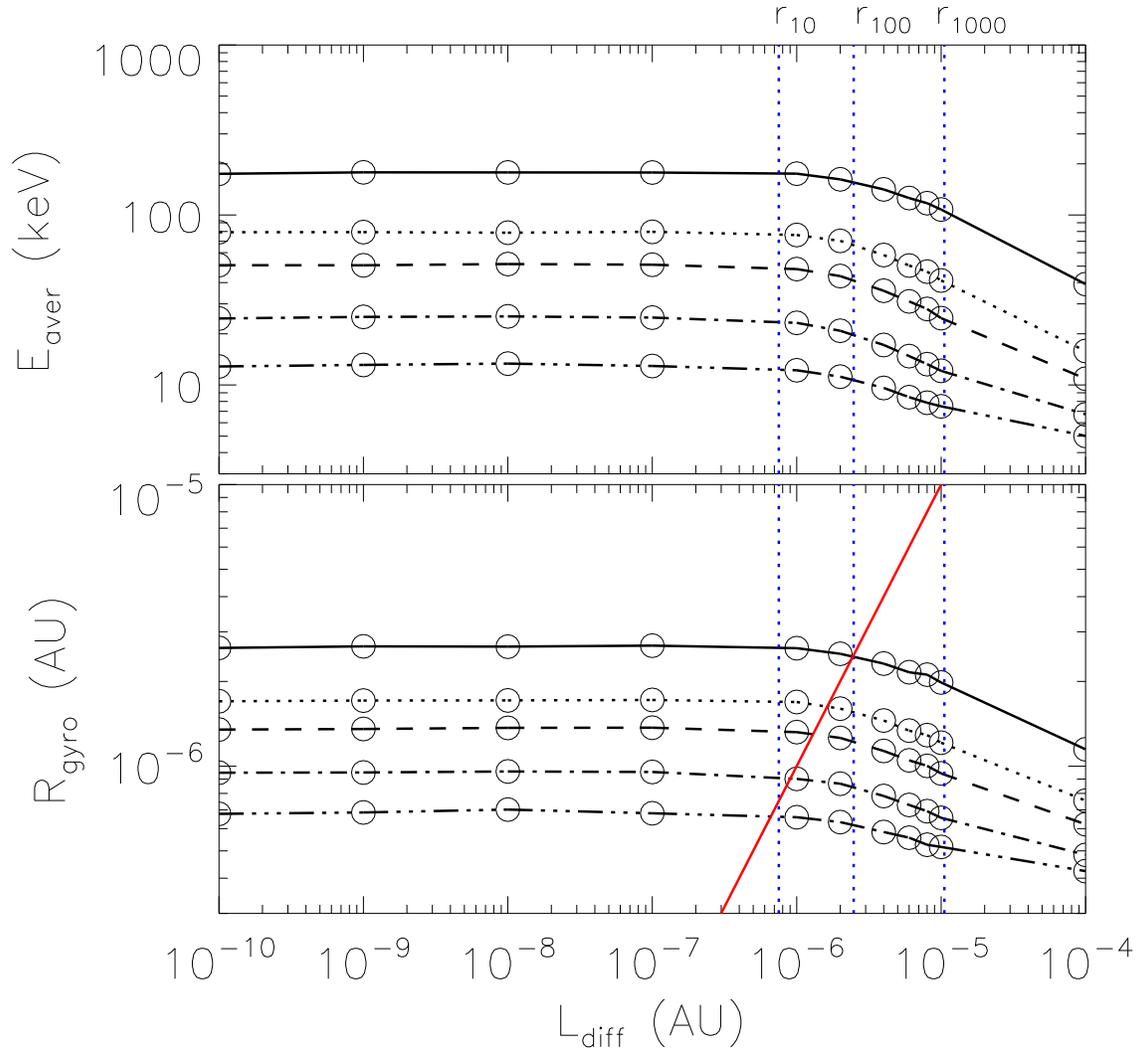}
\caption{Similar to Figure \ref{ldiff0} except that the turbulence level
is $\left(b/B_0\right)^2=1.0$.
\label{ldiff1}}
\end{figure}

\begin{figure}
\epsscale{1.} \plotone{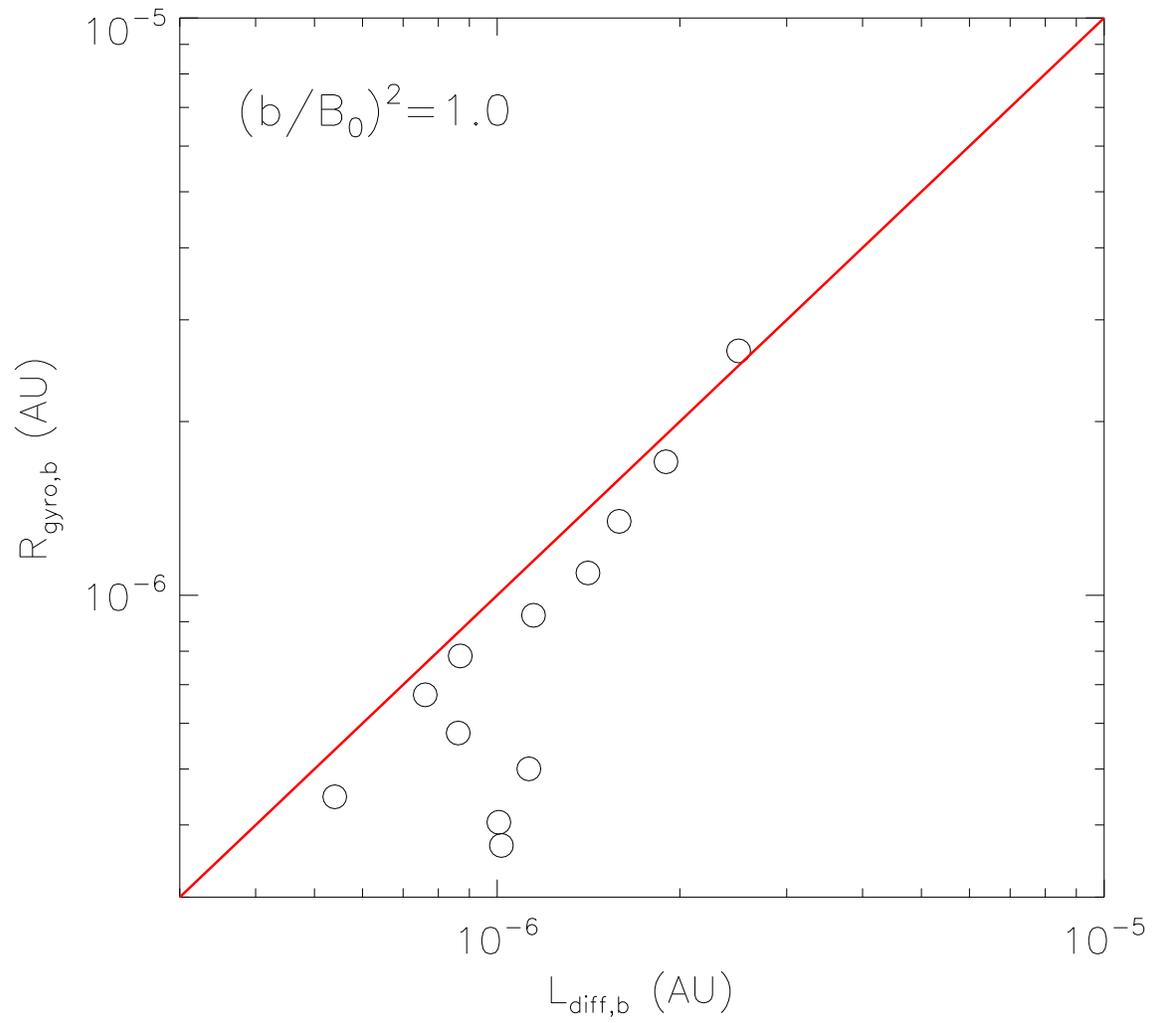}
\caption{Similar as Figure \ref{ldiff0b} except that $\left(b/B_0\right)^2=1.0$.
\label{ldiff1b}}
\end{figure}

Similar phenomenon can also be shown in the case of $\theta_{\text{Bn}}=90^\circ$ 
with higher turbulence level. Figures \ref{ldiff1} and \ref{ldiff1b} are 
similar as Figures \ref{ldiff0} and \ref{ldiff0b}, respectively, except 
that $\left(b/B_0\right)^2=1.0$. Figures \ref{ldiff0}--\ref{ldiff1b} 
suggest that there exists a critical length scale of shock thickness 
for perpendicular shocks in the scale of the average gyro-radii of 
particles with different magnetic turbulence levels $\left(b/B_0\right)^2$, 
the only difference is that with higher $\left(b/B_0\right)^2$ 
particles get weaker acceleration.

\section{CONCLUSIONS AND discussion}

In this paper, we use test particle simulations that include pre-existing 
two-component magnetic field turbulence \citep{QinEA02GRL, QinEA02APJ, 
Qin2002PhDT...1Q} upstream and downstream of the shock \citep{KongEA17} 
to study electron acceleration. In our numerical model, we generate 
magnetic turbulence of slab component with the dissipation range in 
which low-energy electrons resonate, since electrons have higher 
gyrofrequency because of their light mass. We investigate the effects 
of the turbulence level $\left(b/B_0\right)^2$ and shock obliquity 
$\theta_{\text{Bn}}$ on the accelerated electrons spectra.
It is shown that at perpendicular shocks the acceleration of 
electrons, which depends mainly on SDA, is found enhanced under lower 
turbulence levels, since it is suggested that in the presence of turbulence, 
the drift coefficients are reduced \citep[e.g.,][]{EngelbrechtEA17}. 
In addition, at parallel shocks the acceleration of electrons is 
enhanced under higher turbulence levels with the mechanism as the 
following. On the one hand, it is assumed that parallel shocks 
can strongly accelerate particles with first-order Fermi mechanism
if the turbulence level is high since there would exist effective 
particle scatterings to cause multiple shock crossings. On the other
hand, with parallel shocks, particles do not feel drift effects due to 
the large scale background magnetic field, so there is no acceleration 
from SDA in weak turbulence. However, with stronger turbulence, there 
exists large local component of magnetic field perpendicular to the 
shock normal affected by magnetic turbulence, due to which particles 
would feel drift effects, therefore, electrons could get large acceleration from SDA. 
Moreover, for oblique shocks the acceleration of electrons is weak with 
both low and high turbulence levels. Our results also show that parallel 
shock acceleration with large turbulence level (the highest parallel shock 
acceleration) is less effective than perpendicular shock acceleration with 
small turbulence (the highest perpendicular shock acceleration).
The reason might be that SDA is more effective than first-order Fermi 
acceleration, and in a perpendicular shock with low turbulence level, 
the shock acceleration is mainly from SDA, but in a parallel shock with 
high turbulence level, only part of shock acceleration is from SDA.
Recently, \citet{YangEA18} studied acceleration of electrons by ICME-driven 
shocks by comparing strongest parallel and perpendicular shock acceleration 
events observed by the WIND 3DP instrument from 1995 through 2014 at 1 AU. 
They found that quasi-perpendicular shocks are more effective in the 
acceleration of electrons than quasi-parallel shocks are. It is shown that 
our results are compatible to the observations \citep{YangEA18}.
The acceleration efficiency in general increases with the increasing of 
$\theta_{\text{Bn}}$, and its largest variation with $\theta_{\text{Bn}}$ 
is seen at a low turbulence level $\left(b/B_0\right)^2=0.01$. When strong 
magnetic fluctuations, i.e., $\left(b/B_0\right)^2=10.0$, exists at the 
shock front, electron acceleration is found weakly dependent on the 
shock-normal angle, which is in agreement with the study by 
\citet{Guo2015ApJ...802..97G}.
However, although \citet{Guo2015ApJ...802..97G} showed that perpendicular 
shocks with $\left(b/B_0\right)^2=1$ are more effective to accelerate 
particles than that with $\left(b/B_0\right)^2=10$, perpendicular shocks 
with low turbulence level, i.e., $\left(b/B_0\right)^2=0.1$ are less 
effective to accelerate particles than that with high turbulence, i.e., 
$\left(b/B_0\right)^2=1$ and $10.0$. We think the difference is because 
of the difference in the turbulence models we adopt.

Furthermore, we study the impact of shock thickness on the electron 
acceleration at perpendicular shocks. For the dependence of electron 
acceleration on the shock thickness, our simulations with perpendicular 
shocks indicate that there exists a bend-over thickness $L_{\text{diff,b}}$ 
in the scale of particles gyro-radii. The acceleration efficiency does not 
change evidently if the shock thickness is much smaller than $L_{\text{diff,b}}$. 
However, if the shock thickness is much larger than $L_{\text{diff,b}}$, 
the acceleration efficiency starts to drop abruptly. Previous studies have 
shown that the shock thickness may be of the order of ion inertial length 
($c/\omega_{pi}$) \citep{Russell1982GRL...9..1171R, 
Newbury1998JGR...103..29581N}, electron inertial length ($c/\omega_{pe}$) 
\citep{Newbury1996GRL...23..781N, Yang2013PP...20..092116Y},
or convected ion gyroradius ($U_1/\Omega_{ci,2}$) \citep{Bale2003PRL...91..265004B}. 
Different length scale of shock thickness in solar wind would have 
different shock acceleration efficiency. In the condition of this work, 
our simulations show that the bend-over thickness $L_{\text{diff,b}}$ is 
in the order of ion inertial length, however, in other conditions, 
the bend-over thickness may be in other length scale.

In this work, we mainly concentrate on strong shock ($s=4$) acceleration of
electrons. In the
future, we may study the weak strength ($s=2.6$) shock acceleration in detail,
so we are able to investigate what would happen in the termination shock.
In addition, the pickup ion
formation-driven waves which would add an extra component to the slab component
of magnetic turbulence spectrum, 
essentially changing its form and level at wavenumbers corresponding to the
proton gyrofrequency \citep[see, e.g.,][]{Williams1994}, could have a
significant effect on low-energy electron transport parameters in the outer 
heliosphere \citep[e.g.,][]{Engelbrecht2017}.
However, because of our model limitations, we do not consider the pickup ion
formation-driven waves in turbulence. Furthermore, in our turbulence model, we
mainly consider gentle spectral
index ($\beta_d=-2.7$) in the dissipation range. Moreover, we do not include
dissipation range in 2D component. We also do not include the variation of ratio
of slab/2D magnetic turbulence energy in dissipation range. In the future, we 
may modify 
our model to make turbulence more realistic, e.g., we may include
the pickup ion formation-driven waves in our model. We could also consider to
include 2D 
dissipation range in magnetic turbulence. In addition, in dissipation range,
we may vary the turbulence spectra index and the ratio of slab/2D magnetic 
turbulence energy.

\citet{PrinslooEA17} used modulation model to study GCR electrons and
suggested that DSA can explain observed increases in electron intensities at
the termination shock with ad hoc models for the transport 
parameters. It is interesting to find out whether the approach taken in the
numerical simulations in this paper would agree with the conclusions of
\citet{PrinslooEA17} if the model is set up for termination shock conditions
in the future.

\acknowledgments
We gratefully acknowledge Drs. Berndt Klecker, Chao-Sheng Lian, Quanming Lu, and
Yang Wang,  for useful discussions about this topic. This work was supported, in
part, under grant NNSFC 41574172. The work was carried out 
at National Supercomputer Center in Tianjin, and the calculations were performed 
on TianHe-1 (A).

\clearpage

\clearpage

\clearpage

\clearpage

\clearpage

\clearpage

\clearpage

\clearpage

\clearpage

\clearpage
\begin{table}
\caption {Input Parameters for Simulations\label{inputpara}}
\begin{tabular} {ccc}\hline\hline
Parameter && Value \\\hline
$U_1$   &&  500 km s$^{-1}$ \\
$s$     &&  4 \\
$B_{01}$   &&  3 nT \\
$M_{\text{A1}}$ && 10\\
$L_{\text{diff}}$ && 9.28$\times10^{-6}$ AU \\
$\lambda$  && 0.02 AU   \\
$\theta_1$ && 0$^\circ$, 15$^\circ$, 30$^\circ$, 45$^\circ$, 60$^\circ$, 75$^\circ$, 90$^\circ$  \\
$\left(b/B_0\right)^2$ && 0.01, 0.1, 1.0, 10.0    \\
$z_0$    && -5.80$\times 10^{-5}$ AU \\
$E_0$   &&   1 keV          \\
$t_{acc}$ && 23.2 min         \\\hline
\end{tabular}
\end{table}

\end{document}